\documentclass[twocolumn,aps,pr,superscriptaddress,preprintnumbers,nofootinbib,10pt]{revtex4-2}
\usepackage{multirow}
\usepackage{amsmath}
\usepackage{amssymb}
\usepackage[dvipdf,dvips]{graphicx}
\usepackage{color}
\usepackage{hyperref}
\usepackage{url}
\usepackage{slashed}
\usepackage{subfigure}
\usepackage[usenames,dvipsnames]{xcolor}
\usepackage{amsmath}
\usepackage{amsfonts}
\usepackage{float} 
\usepackage{amssymb}
\usepackage{epsfig}
\usepackage{graphics}
\usepackage{euscript}
\usepackage{slashed}
\usepackage{epstopdf}
\usepackage[utf8]{inputenc}
\allowdisplaybreaks
\usepackage[normalem]{ulem}
\usepackage{pifont}
\usepackage{dsfont}
\usepackage{MnSymbol}
\usepackage{verbatim}
\usepackage{graphicx}
\usepackage{latexsym}
\usepackage{tikz-feynman}

\begin{document}

	\title{Numerical approach to the Bell-CHSH inequality in Quantum Field Theory}

	\author{P. De Fabritiis} \email{pdf321@cbpf.br} \affiliation{CBPF $-$ Centro Brasileiro de Pesquisas Físicas, Rua Dr. Xavier Sigaud 150, 22290-180, Rio de Janeiro, Brazil}
	\affiliation{UERJ $–$ Universidade do Estado do Rio de Janeiro,	Instituto de Física $–$ Departamento de Física Teórica $–$ Rua São Francisco Xavier 524, 20550-013, Maracanã, Rio de Janeiro, Brazil}
	
	\author{M. S.  Guimaraes}\email{msguimaraes@uerj.br} \affiliation{UERJ $–$ Universidade do Estado do Rio de Janeiro,	Instituto de Física $–$ Departamento de Física Teórica $–$ Rua São Francisco Xavier 524, 20550-013, Maracanã, Rio de Janeiro, Brazil}
	
	\author{I. Roditi} \email{roditi@cbpf.br} \affiliation{CBPF $-$ Centro Brasileiro de Pesquisas Físicas, Rua Dr. Xavier Sigaud 150, 22290-180, Rio de Janeiro, Brazil}
	
	\author{S. P. Sorella} \email{silvio.sorella@gmail.com} \affiliation{UERJ $–$ Universidade do Estado do Rio de Janeiro,	Instituto de Física $–$ Departamento de Física Teórica $–$ Rua São Francisco Xavier 524, 20550-013, Maracanã, Rio de Janeiro, Brazil}

	\begin{abstract}
		
	The Bell-CHSH (Clauser-Horne-Shimony-Holt) inequality in the vacuum state of a relativistic scalar quantum field is analyzed. Using Weyl operators built with smeared fields localized in the Rindler wedges, the Bell-CHSH inequality is expressed in terms of the Lorentz invariant inner products of test functions. A numerical framework for these inner products is devised. Causality is also explicitly checked by a numerical evaluation of the Pauli-Jordan function. Violations of the Bell-CHSH inequality are reported for different values of the particle mass parameter.
		
	\end{abstract}

	\maketitle

	\section{Introduction}	
	
Entanglement stands, beyond any doubt, as one of the most astounding features of Quantum Mechanics~\cite{Einstein35}, highlighting the existence of remarkably strong correlations between the components of a composite system.  Its importance appears in scales as diverse as that of quantum computing and black hole physics. The Bell-Clauser-Horne-Shimony-Holt (Bell-CHSH) inequality~\cite{Bell64,Clauser69,Clauser74,Clauser78} plays a prominent role in the investigation of this unique property of the quantum world. The violation of this inequality at the quantum level expresses a  profound departure from some of classical physics paradigms, and this has been evidenced by several experiments~\cite{Freedman72,Aspect82,Giustina13,Giustina15,Hensen15, Shalm15, Li18, Rosenfeld17, Storz23}.

Nowadays, there is a significant interest in the investigation of the Bell-CHSH inequalities in the high-energy particle physics context~\cite{Barr24}, enabling to study entanglement at energy scales never explored before. Many experimental tests of the Bell-CHSH inequality in the high-energy regime have been proposed in the last years~\cite{Fabbrichesi21,Afik22,Barr22,Fabbrichesi23, Afik23, Ashby23, Fabbrichesi24}. Quite recently, the first observation of entanglement in a pair of quarks has been reported, being the highest-energy observation of entanglement to date~\cite{Aad23}.
 
Quantum Field Theory (QFT) is the theoretical framework  suitable to handle  properties entailing elementary  particles physics. It looks thus natural to unravel  the Bell-CHSH inequality within the realm of  QFT, to achieve a better understanding of its features in a setting that merges Quantum Mechanics and Special Relativity.
 
Since the pioneering works~\cite{Summers87a,Summers87b,Summers87c}, the study of QFT aspects of the Bell-CHSH has received increasing attention, see \cite{UDW24,Haar23,BRST23,Weyl23,Sinha23,Ghosh23,Morales23} and references therein for recent accounts. 
Facing the Bell-CHSH inequality in QFT forces us to dive  deeper into the complex structure of  quantum fields and  their intrinsic relationship with the Minkowski spacetime. As shown in~\cite{Summers87a,Summers87b}, the Bell-CHSH turns out to be already violated in the vacuum state $\vert 0 \rangle$, when the field operators are localized in the Rindler wedges.

The study of the Bell-CHSH inequality in QFT is rooted  in the so-called Algebraic Quantum Field Theory, making use of the properties of the von Neumann algebras, Tomita-Takesaki modular theory and fundamental theorems such as the Reeh-Schlieder and the Bisognano-Wichmann theorems, crucial tools that only recently are being fully appreciated~\cite{Witten18, Haag92, Bratteli87, Rieffel77, Bisognano:1975ih, Bisognano:1976za}.

Notwithstanding the reach of those studies, most research is centered around its mathematical aspects. Here we are aiming to bridge the gap between the formal approach and a more practical perspective.   We thus provide  a numerical framework to scrutinize the Bell-CHSH inequality in the vacuum state of a  QFT. More precisely, we shall consider the case of a massive scalar  field  in $(1+1)$-dimensional Minkowski spacetime. Upon introducing  the  Lorentz-invariant inner product through the Wightman two-point function for the smeared  fields,  the Bell-CHSH inequality  is stated by employing the unitary Weyl operators and expressed in terms of the inner products between Alice's and Bob's test functions. Selecting the supports of Alice's test functions in the right wedge, we use the Bisognano-Wichmann modular conjugation $j$ \cite{Bisognano:1975ih,Bisognano:1976za}  to obtain  Bob's test functions in the left  wedge, thus ensuring causality. Hence, a numerical setup is devised to evaluate the momentum integrals corresponding to the inner products between test functions and, at the same time,  to check  the vanishing of the Pauli-Jordan function, in order to ensure that causality is correctly implemented.  This enables us to  perform a huge number of  random tests to capture  violations of the  Bell-CHSH inequality for different values of the mass parameter. 
 
This paper is organized as follows. Section~\ref{SecQuantumWeyl} contains a few basic notions about  quantum fields and Weyl operators. In Section~\ref{SecBell} we introduce the Bell-CHSH inequality. The numerical setup as well as the results are presented in Sec.~\ref{SecNumerical}. The conclusion is contained in Sec.~\ref{SecConclusions}.

\newpage

	\section{Quantum fields and Weyl operators}\label{SecQuantumWeyl}
	
	Let us consider a free massive scalar field theory in $(1+1)$-dimensional Minkowski space, with action
	\begin{align} 
		S = \int \!\! d^2x \left[\frac{1}{2} \partial_\mu \phi \, \partial^\mu \phi - \frac{m^2}{2} \phi^2\right].  
	\end{align}
	The scalar field $\phi$ is expanded in terms of creation and annihilation operators as:
	\begin{equation} \label{qf}
		\phi(t,x) = \int \! \frac{d k}{2 \pi} \frac{1}{2 \omega_k} \left( e^{-ik_\mu x^\mu} a_k + e^{ik_\mu x^\mu} a^{\dagger}_k \right), 
	\end{equation} 
	where $\omega_k  = k^0 = \sqrt{k^2 + m^2}$. For the canonical commutation relations one has
	\begin{align}
		[a_k, a^{\dagger}_q] &= 2\pi \, 2\omega_k \, \delta(k - q), \\ \nonumber 
		[a_k, a_q] &= [a^{\dagger}_k, a^{\dagger}_q] = 0. 
	\end{align}
	Quantum fields are operator-valued distributions \cite{Haag92} and have to be smeared to provide well-defined operators acting on the Hilbert space. The smeared quantum field is
	\begin{align} 
		\phi(h) = \int \! d^2x \; \phi(x) h(x),
	\end{align}
	where $h$ is a real smooth test function with compact support, $h \in \mathcal {C}_{0}^{\infty}(\mathbb{R}^4)$. Using the smeared fields, the Lorentz-invariant inner product is introduced by means of the two-point  smeared Wightman function: 
	\begin{align} \label{InnerProduct}
		\langle f \vert g \rangle &= \langle 0 \vert \phi(f) \phi(g) \vert 0 \rangle =  i \Delta(f,g) +  H(f,g)
	\end{align}
	where $ \Delta(f,g)$ and $H(f,g)$ are the smeared versions of the Pauli-Jordan and Hadamard functions:
	\begin{align}
		\Delta(f,g) &=  \int \! d^2x d^2y f(x) \hat{\Delta}(x-y) g(y), \\
		H(f,g) &=  \int \! d^2x d^2y f(x) \hat{H}(x-y) g(y),
	\end{align}
with $\hat{\Delta}(x-y)$ and $\hat{H}(x-y)$ given by
	\begin{align}\label{PJ}
	i \hat{\Delta}(x-y) &= \frac{1}{2} \int \! \frac{dk}{2\pi}  \left(e^{-ik_\mu (x-y)^\mu} - e^{+ik_\mu (x-y)^\mu}\right), \nonumber \\
	\hat{H}(x-y)  &= \frac{1}{2} \int \! \frac{dk}{2\pi}  \left(e^{-ik_\mu (x-y)^\mu} + e^{+ik_\mu (x-y)^\mu}\right).	
\end{align}
We can rewrite the above expressions in a more useful way for our present purposes by moving to momentum space, namely 
\begin{align}
	\langle f \vert g \rangle =  \int \!\! \frac{dk}{2\pi} \frac{1}{2 \omega_k} f^*(\omega_k,k) g(\omega_k,k), \label{aaa}
\end{align}
thus rewriting $\Delta$ and $H$ as
\begin{align}
		i \Delta(f,g) &= \frac{1}{2} \int \! \frac{dk}{2\pi} \frac{1}{2 \omega_k} \left[f^*_k g_k - g^*_k f_k\right], \label{PJ-mom}\\
	H(f,g)  &= \frac{1}{2} \int \! \frac{dk}{2\pi} \frac{1}{2 \omega_k} \left[f^*_k g_k + g^*_k f_k\right], \label{HD-mom}
\end{align}
where $(f_k,g_k)$ stands for $(f(\omega_k,k), g(\omega_k,k))$. 

The Pauli-Jordan distribution is Lorentz-invariant and encodes the information of locality and relativistic causality, being vanishing  outside of the light cone.   Moreover, $\hat{\Delta}(x)$ and $\hat{H}(x)$ are respectively odd and even under the change $x \rightarrow -x$. Using the smeared fields, we can write $\left[\phi(f), \phi(g)\right] = i \Delta(f,g)$, allowing us to recast causality in this setting as having $\left[\phi(f), \phi(g)\right] = 0$ when the supports of $f$ and $g$ are spacelike separated.
	
	Let us introduce the unitary Weyl operators \cite{Weyl23}, defined as 
	\begin{align}
		A_f = e^{i \phi(f)}.
	\end{align}
	These operators obey the so-called Weyl algebra: 
	\begin{align}\label{WeylAlgebra}
		A_f \,  A_g = e^{-\frac{i}{2} \Delta(f,g)} \,A_{(f+g)}.
	\end{align}
	When the supports of $f$ and $g$ are spacelike separated, the Pauli-Jordan function vanishes, ensuring that $A_f A_g = A_{(f+g)}$.   Computing the vacuum expectation value of the Weyl operator, one finds:
	\begin{align}\label{WeylVEV}
		\langle 0 \vert A_f \vert 0 \rangle = \langle 0 \vert A_{-f} \vert 0 \rangle = e^{-\frac{1}{2} \vert\vert f \vert\vert^2},
	\end{align}
	where  $\vert\vert f \vert\vert^2 = \langle f \vert f \rangle$.

\section{The Bell-CHSH inequality}\label{SecBell}
	
	We proceed now by stating the Bell-CHSH inequality in the vacuum state of the scalar field. We shall follow the procedure outlined in~\cite{Guimaraes:2024alk}, where the Bell-CHSH correlator has been formulated in terms of unitary operators. To that purpose one considers an open region $O$ in Minkowski space and define $\mathcal{M}(O)$ as the space of smooth test functions with compact support contained in $O$. One introduces  its complement $\mathcal{M}'(O)$ as the set of test functions that have a vanishing smeared Pauli-Jordan function with all test functions in $\mathcal{M}(O)$, that is, $\mathcal{M}'(O) = \{g; \Delta(f,g)=0, \forall f \in \mathcal{M}(O)\}$. Therefore, one rephrases causality as
	\begin{align}
		\left[\phi(f), \phi(g)\right] = 0, \quad \forall f \in \mathcal{M}(O) \,\, {\rm and} \,\, \forall g \in \mathcal{M}'(O).
	\end{align}
	
	Let us introduce Alice's operators $(A_f, A_{f'})$ as the Weyl operators $A_f = e^{i \phi(f)}$ and $A_{f'} = e^{i \phi(f')}$ built with test functions $f, f' \in \mathcal{M}(O)$.  We introduce Bob's operators $(A_g, A_{g'})$ as $A_g = e^{i \phi(g)}$ and $A_{g'} = e^{i \phi(g')}$, built with test functions $g, g' \in \mathcal{M}'(O)$, thus guaranteeing that $\left[A_f, A_g\right]=\left[A_f, A_{g'}\right]=\left[A_{f'}, A_g\right]=\left[A_{f'}, A_{g'}\right]=0$ since the Pauli-Jordan function vanishes for these test functions. As such,  the corresponding Weyl operators commute.  
	
	Following~\cite{Guimaraes:2024alk}, the Bell-CHSH correlation function in the vacuum state reads
	\begin{align}
		\langle 0 \vert \mathcal{C} \vert 0 \rangle &= 	\langle 0 \vert \left(A_f + A_{f'}\right) A_g + \left(A_f - A_{f'}\right) A_{g'} \vert 0 \rangle.
	\end{align}
	One speaks of a violation whenever 
	\begin{equation} 
	2 < | \langle 0 \vert \mathcal{C} \vert 0 \rangle | \le 2 \sqrt{2}. 
	\end{equation}
	From  Eqs.~\eqref{WeylAlgebra} and~\eqref{WeylVEV}, it follows 
	\begin{align}
		\langle 0 \vert A_f A_g \vert 0 \rangle = 	\langle 0 \vert A_{(f+g)} \vert 0 \rangle = e^{-\frac{1}{2} \vert\vert f+g \vert\vert^2}.
	\end{align} 
	Therefore, for the Bell-CHSH correlator defined above, one has
	\begin{align}\label{BellFGpre}
		\langle 0 \vert \mathcal{C} \vert 0 \rangle &= e^{-\frac{1}{2} \vert\vert f + g \vert\vert^2} + e^{-\frac{1}{2} \vert\vert f' + g \vert\vert^2} \nonumber \\
		&+ e^{-\frac{1}{2} \vert\vert f + g' \vert\vert^2} - e^{-\frac{1}{2} \vert\vert f' + g' \vert\vert^2}
	\end{align}
	To evaluate the norms appearing in Eq.~\eqref{BellFGpre}, we remind that 
 from Eq.~\eqref{InnerProduct}, one has  $\langle f \vert g \rangle = i\Delta(f,g) + H(f,g)$. Since $f \in \mathcal{M}$ and $g \in \mathcal{M}'$, one finds $\Delta(f,g)=0$. Also, for any test function $f$, it holds that $\Delta(f,f)=0$ as can be seen from Eq.~\eqref{PJ-mom}. Therefore	
	 \begin{align}\label{NormaFG}
			\vert\vert f+g \vert\vert^2 = H(f,f) + 2 H(f,g) + H(g,g).
	\end{align}
	Putting together Eqs.~\eqref{BellFGpre} and~\eqref{NormaFG}, one finds 
	\begin{align}\label{BellFG}
		\langle 0 \vert \mathcal{C} \vert 0 \rangle &= e^{-\frac{1}{2}\left[H(f,f) + 2 H(f,g) + H(g,g)\right]} \nonumber \\
		&+ e^{-\frac{1}{2} \left[H(f',f') + 2 H(f',g) + H(g,g)\right]} \nonumber \\
		&+ e^{-\frac{1}{2} \left[H(f,f) + 2 H(f,g') + H(g',g')\right]}  \nonumber \\
		&- e^{-\frac{1}{2} \left[H(f',f') + 2 H(f',g') + H(g',g')\right]}.
	\end{align}
	
	Expression~\eqref{BellFG} is the starting point for the numerical setup which will be outlined in the following. Though, besides the evaluation of the Hadamard smeared functions present in Eq.~\eqref{BellFG}, we shall also check the Pauli-Jordan factors $\Delta(f,g)$, $\Delta(f',g)$, $\Delta(f,g')$, $\Delta(f',g')$ to ensure that causality is numerically fulfilled. 

		Let us conclude this section by underlining that, as already mentioned, the region ${\cal O}$ and its causal complement ${\cal O}'$ will be identified with the right and left Rindler wedges, namely 
		\begin{equation} 
		{\cal W}_{R} =  \{\; x; \; x> |t| \; \} \;, \qquad {\cal W}_{L} =  \{\; x; \; -x> |t| \; \} \;. \label{Rindler}
		\end{equation}

	\section{Numerical setup}\label{SecNumerical}

	Let us define the following Gaussian functions:
	\begin{align}
		\varphi\left(t,x\right) &= \alpha \, e^{-\frac{1}{2}\frac{\left(t-t_0\right)^2}{\delta t^2}} \, e^{-\frac{1}{2}\frac{\left(x-x_0\right)^2}{\delta x^2}}, \nonumber \\
		\psi\left(t,x\right) &= \beta \, e^{-\frac{1}{2}\frac{\left(\tau - \tau_0\right)^2}{\delta \tau^2}} \, e^{-\frac{1}{2}\frac{\left(y - y_0\right)^2}{\delta y^2}},
	\end{align}
	where $\alpha$ and $\beta$ are arbitrary real numbers.
	Notice that $\varphi$ and $\psi$ depends on the parameters $\{t_0, \delta t, x_0, \delta x\}$ and $\{\tau_0, \delta \tau, y_0, \delta y\}$. These parameters determine the locations and the widths of the Gaussians in Minkowski space. Their Fourier transforms are
	\begin{align}
	\varphi\left(p\right) &= \alpha \, 2 \pi \delta t \delta x \, e^{-\frac{1}{2}(\omega_p^2 \delta t^2 + p^2 \delta x^2)} e^{i \omega_p t_0 + i p x_0} \nonumber \\
	\psi\left(p\right) &= \beta \, 2 \pi \delta \tau \delta y \, e^{-\frac{1}{2}(\omega_p^2 \delta \tau^2 + p^2 \delta y^2)} e^{i \omega_p \tau_0 + i p y_0},
	\end{align}
	where we will later impose $\omega_p = \sqrt{p^2 + m^2}$. 
	
	Gaussians are smooth but do not have compact support, being in principle not appropriate to be used as test functions. Nevertheless, this issue can be handled with the help of the Pauli-Jordan function. To illustrate this point, let $(h_{R}, h_{L})$ stand for two generic  Gaussians located in  ${\cal W}_R$ and in  ${\cal W}_L$, respectively. The first observation relies on the possibility of shaping $(h_{R}, h_{L})$ by a convenient choice of the widths, in such a way that they resemble very much compactly supported functions, meaning that they decay very fast outside spacetime regions located in ${\cal W}_R$ and ${\cal W}_L$.  Doing so, one can check if $(h_{R}, h_{L})$ can be effectively considered to be spacelike by looking at the Pauli-Jordan function $\Delta(h_{R}, h_{L})$ and demanding that. 
\begin{equation} 
\Delta(h_{R}, h_{L}) \le 10^{-10} \;. \label{ten}
\end{equation}
In other words, Gaussian functions for which $\Delta(h_{R}, h_{L}) > 10^{-10}$ will be automatically rejected. Only  when $\Delta(h_{R}, h_{L}) \le 10^{-10}$, they will be employed to probe the violation of the Bell-CHSH inequality. 	The numerical threshold \eqref{ten} has been established after performing a huge number of tests, typically half a million tests for each value of the mass parameter.  Observe that the value $10^{-10}$ can be, in practice,  considered as vanishing when  compared with the size of the violations  reported in Table \eqref{tabelaMax}. 
	
Let us proceed by specifying how Bob's Gaussians in ${\cal W}_L$ are obtained from Alice's Gaussians in ${\cal W}_R$. To that aim we employ the Bisognano-Wichmann \cite{Bisognano:1975ih,Bisognano:1976za} modular conjugation $j$, whose action on the test functions is defined by 
	\begin{align}
		j \varphi(t, x) = \varphi(-t, -x), \quad  j \psi(t, x) = \psi(-t, -x).
	\end{align}
	Notice that this transformation for $\varphi$ is equivalent to changing $(t_0, x_0) \rightarrow (-t_0, -x_0)$. The same holds  for  $\psi$. Needless to say, the modular conjugation $j$ is nothing but the $CPT$ operator \cite{Bisognano:1975ih,Bisognano:1976za}. A Gaussian  centered at $(t_0, x_0)$ is transformed into a Gaussian centered at  $(-t_0, -x_0)$. Thus, if $\varphi$ is located in  ${\cal W}_R$, under the action of $j$, it will be transformed in a Gaussian located in ${\cal W}_L$. 
	
	In the following, we will consider $\varphi$ and $\psi$ as Alice's test functions $f, f'$. Correspondingly,  $j \varphi$ and $j \psi$ are Bob's test functions $g, g'$. We will consider $\varphi$ and $\psi$ centered in the upper part of ${\cal W}_R$, that is, we will take $t_0, x_0, \tau_0, y_0 > 0$, $x_0 > t_0$ and $y_0 > \tau_0$. As such,  $j\varphi$ and $j\phi$ turn out to be  centered in ${\cal W}_L$. Although their centers are in spacelike separated regions, because of their widths, a careful check of the vanishing of the relevant Pauli-Jordan functions will be performed.  This will enable to use Eq.~\eqref{NormaFG}.
	
	The steps of the method adopted can be resumed as follows. We randomly choose points $(t_0, x_0)$ and $(\tau_0, y_0)$ in the upper part of ${\cal W}_R$, as well as real positive values for $\alpha$ and $\beta$, sampling these parameters in the range $(0, 10)$ using the {\it Mathematica} command ``RandomReal". We also fix the widths $\delta t = \delta x = \delta \tau = \delta y = 0.1$ for definiteness. We now pick up  a  value for the mass $m$ and evaluate all the Hadamard functions needed for the Bell-CHSH inequality~\eqref{BellFG}. Next, with the same set of parameters, we test all the relevant Pauli-Jordan functions, in order to check if causality is fulfilled, according to the threshold~\eqref{ten}. All integrals are estimated numerically using the software {\it Mathematica}\footnote{We mainly used method {\it DoubleExponentialOscillatory}, but we also checked the consistency of the numerical integration by adopting other methods.}.  
	Then, we check if for this set of parameters a violation of the Bell-CHSH inequality is found, that is, if $\lvert \langle 0 \vert \mathcal{C} \vert 0 \rangle \lvert > 2$.  We repeat then the procedure $5 \times 10^5$ times for each mass value\footnote{To give an idea about the computing time on a single  laptop, each mass value required around 15 days to run the tests.}.

	Six different values of the mass have been considered: $m=1.5,1, 0.1, 0.01, 0.001, \, {\rm and} \, 0.0001$. The maximum violations achieved for each case are reported in Table~\ref{tabelaMax}. 	
	\begin{table}[h!]
		\begin{tabular}[t]{|c|c|}
			\hline
			$m$ &
			$\langle \mathcal{C} \rangle$  \\
			\hline 
			1.5 & 2.00148 \\
			\hline
			1 & 2.00722  \\
			\hline
			0.1 & 2.06382 \\
			\hline
			0.01 & 2.10044 \\
			\hline
			0.001 & 2.12661 \\
			\hline
			0.0001 & 2.13046 \\
			\hline
		\end{tabular}
		\caption{Maximum violation achieved for each mass value. For the corresponding Pauli-Jordan functions, we always have $\Delta \le 10^{-10}$.}
		\label{tabelaMax}
	\end{table}	
	
One notices that the size of the violation decreases with the increase of the mass $m$. This feature corroborates the observation of \cite{Summers87a}, according to which, due to the cluster property of a massive QFT\footnote{Meaning that sufficiently distant regions of a quantum field system become statistically independent.}, the violation of the Bell-CHSH is expected to decay exponentially with the spatial separation between Alice's and Bob's region. We note in fact that, for large spatial separation $L$, the Hadamard function behaves as\\
	\begin{equation} 
	H(0,L) = \int \frac{dk}{2\pi} \frac{1}{2 \omega_k} \cos(kL) \approx e^{-m L} \;. \label{cluster} 
	\end{equation}
This equation provides a simple understanding of the fact that using Weyl operators yields a Bell-CHSH inequality violation in agreement with the cluster property.

	\section{Conclusions}\label{SecConclusions}
	
	In this work we developed a numerical setup for studying the Bell-CHSH inequality in the vacuum state of a relativistic scalar massive  Quantum Field Theory in $(1+1)$ Minkowski spacetime. We defined a Lorentz-invariant inner product through the Wightman smeared two-point function and introduced the Bell-CHSH inequality by means of Weyl operators. The integrals present in the terms of the inequality were computed numerically and we devised a random test in order to search for parameters leading to Bell-CHSH inequality violations for different values of the particle mass, see Table \eqref{tabelaMax}.

	Let us end by elaborating on the following topics, all under current investigation:
	\begin{itemize}
	\item As already mentioned, in the case of a relativistic scalar field, several results have been achieved on the violation of the Bell-CHSH inequality in the vacuum state~\cite{Summers87a,Summers87b,Weyl23}, relying on a set of mathematical tools such as von Neumann algebras, Tomita-Takesaki modular theory and the properties of  Weyl operators. Those results are of a theoretical nature, aiming at establishing general features of the Bell-CHSH inequality. From this perspective, the construction of an explicit numerical setup is  welcome. To our knowledge, this is the first fully numerical attempt to face the Bell-CHSH inequality for a scalar field, including the handling of the check of causality through the Pauli-Jordan function. At the present stage, besides a few technical points, we do not see major obstacles to generalize the present framework to the case of a scalar field in $1+3$ Minkowski spacetime as well as to the case of spinor fields. A point which deserves a certain care is the fact that, in $1+3$ dimensions, the scalar product, Eq.~\eqref{aaa}, would read
	\begin{eqnarray}
	\langle f \vert g \rangle &  = &  \int \!\! \frac{d^3k}{(2\pi)^3} \frac{1}{2 \omega_k} f^*(\omega_k,{\vec k}) g(\omega_k,{\vec k}), \nonumber \\
	& =&  \int \!\! \frac{ {k^2 } \sin{\theta} dk \;d \theta \; d \varphi}{2 (2\pi)^3  \omega_k} \;f^*g
	\label{aa1}
\end{eqnarray}
where
\begin{align} 
f= f(\sqrt{k^2+m^2},&\; k \sin{\theta} \cos{\varphi}, \nonumber \\&\,  k \sin{\theta} \sin{\varphi},\, k \cos{\theta}), \label{aa2}
\end{align}
and similarly for $g$. The above expression exhibits the challenges of the $1+3$ case. Multiple integrals need to be evaluated numerically, a feature which might require a number of additional tools. This issue is under consideration and will be reported as soon as possible.

\item As shown in Table~\eqref{tabelaMax}, the size of the violation decreases with the increase of the mass $m$. This result is deeply related to the cluster property, a general feature of the correlation functions of Quantum Field Theory. Due to its relevance, it is worth to provide here a more detailed discussion. As far as the Bell-CHSH inequality is concerned, we can distinguish basically two cases. The first one is that faced in the present work. More precisely, the test functions $f$ and $g$ refer to two regions, $O$ and $O'$, located, respectively, in the right and left Rindler wedges, $({\cal W}_R, {\cal W}_L)$, Eq.~\eqref{Rindler}. Essentially, $O$ and $O'$ are identified with the two tiny regions in which $f$ and $g$ are effectively nonvanishing, Eq.~\eqref{ten}. In this case $O$ and $O'$ are not the causal complement of each other. A spatial separation between them can be easily visualized. In such a situation, the Bell-CHSH is expected to display a decay with the increase of the mass, in agreement with the cluster property, see~\cite{Summers87a}. This is exactly what has been detected: a clear dependence from the mass parameter $m$. There is, however, a second situation which requires a fully different numerical handling. It is the case in which the two regions $O$ and $O'$ coincides with the whole wedges $({\cal W}_R, {\cal W}_L)$. This means that the test functions $f$ and $g$ are smooth functions supported in $({\cal W}_R, {\cal W}_L)$, which vanish outside of $({\cal W}_R, {\cal W}_L)$. An example of such a test functions is given, for instance, by
\begin{align}
f(t,x) = 
\left\{
    \begin {aligned}
         & e^{-\frac{1}{x^2-t^2}}\;e^{-\frac{x^2}{2}} \quad & x \geq |t|  \\
         & 0 \quad & {\rm elsewhere}                   
    \end{aligned}
\right. \label{aa3}
\end{align}
where the exponential factor $e^{-\frac{x^2}{2}}$ ensures the needed decaying behavior at $x \rightarrow \infty$. The function $f$ in expression~\eqref{aa3} is smooth and defined in the whole right wedge ${\cal W}_R$, which vanishes at the boundary, {\it i.e.} $t=\pm x$. Analogously, in the left wedge ${\cal W}_L$, one might consider
\begin{align}
g(t,y) = 
\left\{
    \begin {aligned}
         & e^{-\frac{1}{y^2-t^2}}\;e^{-\frac{y^2}{2}} \quad & -y \geq |t|  \\
         & 0 \quad & {\rm elsewhere}                   
    \end{aligned}
\right. \label{aa4}
\end{align}

This situation is different from the previous one. Here, we are dealing with two regions,  $({\cal W}_R, {\cal W}_L)$, which are the causal complement of each other. Moreover, there is no spatial separation between them as they touch at the origin, $x=t=0$. For such a spacetime configuration, the theoretical results~\cite{Summers87a,Summers87b} predict that the violation of the Bell-CHSH inequality achieves its maximum value, namely $2\sqrt{2}$. Unfortunately, the analytic expression of the Fourier transformation of expressions~\eqref{aa3} and~\eqref{aa4} is not available. As a consequence, a new different numerical setup would be needed. Here, we limit ourselves to mention that we are trying to figure out if a possible framework based on configuration space and not on momentum space might be viable.
	
From this discussion, the various numerical aspects and challenges related to the issue of the cluster property should become more apparent. In particular, the geometrical configuration faced in the present work exhibits a clear connection with the cluster property, as evidenced by the results of Table~\eqref{tabelaMax}.
\end{itemize} 

\vspace{1cm}

	\section*{Acknowledgments}
	The authors would like to thank the Brazilian agencies CNPq and CAPES, for financial support.  S.P.~Sorella, I.~Roditi, and M.S.~Guimaraes are CNPq researchers under contracts 301030/2019-7, 311876/2021-8, and 309793/2023-8, respectively. P.~De~Fabritiis acknowledges FAPERJ for financial support	under Contract No. SEI-260003/000133/2024.


		
\end{document}